\documentclass[Afour,times,sagev]{sagej}

\usepackage{moreverb,url}
\usepackage{mathabx}
\usepackage[ruled,linesnumbered]{algorithm2e}
\usepackage{bm}

\usepackage[colorlinks,bookmarksopen,bookmarksnumbered,citecolor=blue,urlcolor=blue,linkcolor = blue]{hyperref}

\newcommand\BibTeX{{\rmfamily B\kern-.05em \textsc{i\kern-.025em b}\kern-.08em
T\kern-.1667em\lower.7ex\hbox{E}\kern-.125emX}}

\def\volumeyear{2023}

\begin{document}


\title{Estimating the EVSI with Gaussian Approximations and Spline-Based Series Methods}

\author{Linke Li\affilnum{1,2}, Hawre Jalal\affilnum{3}, and Anna Heath\affilnum{1,2,4}}

\affiliation{\affilnum{1}Dalla Lana School of Public Health, University of Toronto, Toronto, Canada; \affilnum{2}Child Health Evaluative Sciences, The Hospital for Sick Children, Toronto, Canada; \affilnum{3}School of Epidemiology and Public Health, University of Ottawa, Ottawa, Canada; \affilnum{4}Department of Statistical Science, University College London, London, United Kingdom;}



\begin{abstract}

\textbf{Background}. The Expected Value of Sample Information (EVSI) measures the expected benefits that could be obtained by collecting additional data. Estimating EVSI using the traditional nested Monte Carlo method is computationally expensive but the recently developed Gaussian approximation (GA) approach can efficiently estimate EVSI across different sample sizes. However, the conventional GA may result in biased EVSI estimates if the decision models are highly nonlinear. This bias may lead to suboptimal study designs when GA is used to optimize the value of different studies. Therefore, we extend the conventional GA approach to improve its performance for nonlinear decision models.
\textbf{Methods}. Our method provides accurate EVSI estimates by approximating the conditional expectation of the benefit based on two steps. First, a Taylor series approximation is applied to estimate the conditional expectation of the benefit as a function of the conditional moments of the parameters  of interest using a spline, which is fitted to the samples of the parameters and the corresponding benefits. Next, the conditional moments of parameters are approximated by the conventional GA and Fisher information. The proposed approach is applied to several data collection exercises involving non-Gaussian parameters and nonlinear decision models. Its performance is compared with the nested Monte Carlo method, the conventional GA approach, and the nonparametric regression-based method for EVSI calculation.
\textbf{Results}. The proposed approach provides accurate EVSI estimates across different sample sizes when the parameters of interest are non-Gaussian and the decision models are nonlinear. The computational cost of the proposed method is similar to other novel methods.
\textbf{Conclusions}. The proposed approach can estimate EVSI across sample sizes accurately and efficiently, which may support researchers in determining an economically optimal study design using EVSI.

\end{abstract}

\Highlights{1. The Gaussian approximation method efficiently estimates the Expected Value of Sample Information for clinical trials with varying sample sizes, but it may introduce bias when health economic models have nonlinear structure.\\
2. We introduce the spline-based Taylor series approximation method and combine it with the original Gaussian approximation to correct the nonlinearity-induced bias in EVSI estimation.\\
3.  Our approach can provide more precise EVSI estimates for complex decision models without sacrificing computational efficiency, which can enhance the resource allocation strategies from the cost-effective perspective.}



\maketitle

\section{INTRODUCTION}
Value of information (VoI) analysis involves calculating the economic benefits of reducing uncertainty in a decision model.\cite{schlaifer1967applied,howard1966information,pratt1995introduction,briggs2006decision} In general, VoI methods use Bayesian decision theory to integrate insights from health economic decision models with data from previous studies. This approach aids in informing decision-making regarding future data collection efforts and resource allocation. \cite{schlaifer1967applied,jackson2021value} One of the metrics in VoI analysis involves the Expected Value of Sample Information (EVSI) which quantifies the expected economic benefits obtained from a specific data collection experiment. EVSI has a high potential to assist in selecting the optimal trial from a health economic standpoint and can guide public resource allocation to optimize future data collection efforts.\cite{willan2005value,kunst2020computing,fairley2020optimal,claxton2005probabilistic,canadian2006guidelines,flight2021expected}

However, computing EVSI has been traditionally challenging due to its conceptual and computational burdens.  This is because EVSI requires integrating over the conditional expectation of the net benefits, which is usually estimated through nested simulations.\cite{ades2004expected} Most simulation models used in practice are complex without the possibility of analytically computing the conditional expectation.  Thus, EVSI is often computed numerically by estimating the conditional expectation of the net benefits a large number of times. Traditional estimation methods involve numerical methods such as Markov Chain Monte Carlo (MCMC) and may need considerable time to evaluate.\cite{ades2004expected}  As a result, the computation burden of EVSI is often high especially when the health economic decision model is complex. This has limited the application of EVSI in practice.\cite{heath2020calculating,steuten2013systematic} 

In recent years, several methods have been proposed to reduce the computational cost of estimating EVSI. \cite{brennan2007efficient,kharroubi2011estimating,strong2015estimating,menzies2016efficient,hironaka2020multilevel,heath2018efficient,heath2019estimating,jalal2018gaussian}
These methods have been applied in real-world studies. \cite{kunst2020computing, heath2020calculating,kunst2019value}  In this exposition, we will be using the Gaussian approximation (GA) approach, which can easily identify the optimal design as it estimates EVSI for studies with distinct sample sizes at a low computational cost.\cite{jalal2018gaussian}  Specifically, the GA approach is based on approximating the conditional expectation of net benefit for each intervention by fitting a model between samples of net benefits and parameters of a health economic decision model.  \cite{jalal2018gaussian,doubilet1985probabilistic}   However, the GA approach may produce biased estimates of EVSI when dealing with nonlinear models. This is because the GA approach estimates the conditional expectation of the net benefit using an estimate of the conditional expectation of the model parameters. Such a process may inadequately estimate the conditional expectation of the net benefit when the relationship between the parameters and the net benefit is non-linear. \cite{jalal2018gaussian} 

To improve the EVSI estimation accuracy of GA for nonlinear decision models, we extend the original GA by introducing higher-order correction terms. This is achieved by decomposing the conditional expectation of net benefits into two components using Taylor series expansions.\cite{wolter2007introduction,oehlert1992note,wang2021package,wang2021shape} The first component is estimated using the original GA approach, while the second component is estimated using the fitted model, samples of parameters, and expected Fisher information to adjust for bias resulting from nonlinear net benefits functions.\cite{bernardo2009bayesian, van2000asymptotic, gelman2013bayesian} The resulting method is named the spline-based Taylor series approximation and Gaussian approximation (TGA). This approach retains its efficiency in approximating EVSI across sample sizes once the prior effective sample sizes are obtained.

We begin this article by formally introducing the definition of EVSI and reviewing the methodology of the GA approach. \cite{jalal2018gaussian} Then, we present our extension to the GA methods by first presenting how to approximate the conditional expectation of the net benefits using Taylor series expansions, which is followed by an introduction to the approximation methods of the elements required for conducting Taylor series approximation. After that, we use stylized examples and the case study model included in the original GA article to demonstrate that our extended GA approach can provide more accurate EVSI estimates for nonlinear decision models. \cite{jalal2018gaussian} We conclude this paper with a short discussion.

\section{METHODS}

\subsection{The Expected Value of Sample Information}

Health economic decision models assess the net monetary (or net health) benefits of different interventions to help decision-makers choose the optimal decision from $D$ competing alternatives. \cite{stinnett1998net} Denoting the parameters in this health economic decision model by $\bm{\theta}$, which includes all parameters used to compute the benefit of each intervention, the population-level net benefit for each decision $d = 1, \dots, D$ is defined as the difference between the decision monetary value of this decision -- as reflected by the underlying health economic decision model -- and its costs. The net benefit of the decision $d$ is a function of $\bm{\theta}$ and is denoted by $\mbox{NB}_d(\bm{\theta})$. The current knowledge about model parameters is summarized as a joint probabilistic distribution $p(\bm{\theta})$, which then induces uncertainty in the net benefit. Note that, $p(\bm{\theta})$ can also be considered as the prior distribution of $\bm{\theta}$ in Bayesian statistics. The statistical uncertainty in the net benefits and model parameters is usually explored through a set of simulations with $M$ samples of $\bm{\theta}$ and the corresponding samples of $\mbox{NB}_d(\bm{\theta}), d = 1, \dots, D$. This dataset is known as the probabilistic analysis (PA) dataset.

Using the information contained in the health economic decision model, the Expected Value of Sample Information (EVSI) is defined as the difference between the expected net benefit provided by the optimal decision that is made after an additional dataset is collected, versus the expected benefit given by the optimal decision based on prior knowledge. Using our prior knowledge about  $\bm{\theta}$, the expected benefit given by the optimal decision is:

\begin{equation}
    \max_d \mathbb{E_{\bm{\theta}}}\{ \mbox{NB}_d(\bm{\theta}) \} . \label{eq:EVSI_prior}
\end{equation}

For a study with a sample size $n$, we plan to collect a dataset $\bm{X}_n = (X_1, \dots, X_n)$  to inform the parameters $\bm{\theta}$. $\bm{X}_n$ can be informative for all the model parameters $\bm{\theta}$, or only informative for a subset of $\bm{\theta}$, $\bm\phi$. If $\bm{X}_n$ had been collected, we would then compute the conditional distribution of the parameters $p(\bm{\theta}|\bm{X}_n)$, which would update the distribution of the net benefits and may change the optimal decision. The expected  benefit of the optimal decision based on the updated parameter distribution is defined as
\begin{equation*}
    \max_d  \mathbb{E}_{\bm{\theta}}[ \mbox{NB}_d(\bm{\theta})|{\bm{X}_n}].
\end{equation*}

As these data have not been collected, EVSI is defined by averaging over all potential datasets. The probabilistic distribution of potential datasets $\bm{X}_n$ is determined by both the likelihood function $p(\bm{X}_n|\bm{\theta})$ and the prior distribution $p(\bm{\theta})$:
\begin{equation}
    p(\bm{X}_n) = \int_{\bm{\theta}} p(\bm{\theta})p(\bm{X}_n|\bm{\theta}) d \bm{\theta}.\label{eq:marginal}
\end{equation}

We then average over the randomness of $\bm{X}_n$ to assess the total expected benefit that can be obtained by collecting an additional dataset:
\begin{equation}
\mathbb{E}_{\bm{X}_n} \left [ \max_d \mathbb{E}_{\bm{\theta}}[ \mbox{NB}_d(\bm{\theta}) |{\bm{X}_n}] \right ].\label{eq:EVSI_posterior}
\end{equation}

EVSI for a study design with a sample size $n$ is defined as the difference between terms \ref{eq:EVSI_posterior} and \ref{eq:EVSI_prior}: 
\begin{align}
    \text{EVSI}(n) &= \mathbb{E}_{\bm{X}_n} \left [ \max_d \mathbb{E}_{\bm{\theta}}[ \mbox{NB}_d(\bm{\theta}) |{\bm{X}_n}] \right ] \notag \\ 
    & -  \max_d \mathbb{E}_{\bm{\theta}}\{ \mbox{NB}_d(\bm{\theta}) \} \label{eq:EVSI_fixed_2} \\
 &= \mathbb{E}_{\bm{X}_n} \left [ \max_d \mathbb{E}_{\bm{\theta}}[ \mbox{NB}_d(\bm{\theta}) |{\bm{X}_n}] \right ] \notag \\ 
    & -  \max_d \mathbb{E}_{\bm{X}_n}
    \left[ \mathbb{E}_{\bm{\theta}}[ \mbox{NB}_d(\bm{\theta}) |{\bm{X}_n}]  \right], \label{eq:EVSI_fixed}
\end{align}
where  equations \ref{eq:EVSI_fixed_2} and \ref{eq:EVSI_fixed} are equivalent because of the law of total expectation. \cite{strong2015estimating} Equation \ref{eq:EVSI_fixed} is more commonly used in numerical approximation of EVSI since it can reduce the uncertainty introduced by Monte Carlo sampling.  \cite{strong2015estimating} 

The traditional method for calculating the first component in Equation \ref{eq:EVSI_fixed} is Monte Carlo sampling, which requires a nested 2-stage process: first, we need to simulate a large number of samples of datasets $\bm{X}_n$ using  equation \ref{eq:marginal}. Then, for each dataset, we generate the conditional distribution $p(\bm{\theta}|\bm{X}_n)$ and compute the maximized averaged net benefit $\max_d \mathbb{E}_{\bm{\theta}}[ \mbox{NB}_d(\bm{\theta}) |{\bm{X}_n}]$. \cite{heath2022simulating} After that, we average  $\max_d \mathbb{E}_{\bm{\theta}}[ \mbox{NB}_d(\bm{\theta}) |{\bm{X}_n}]$ over the samples of simulated datasets $\bm{X}_n$ to approximate the outer expectation.\cite{ades2004expected} Since  $p(\bm{\theta}|\bm{X}_n)$ does not always have a closed-form solution, an MCMC method may be required to approximate $p(\bm{\theta}|\bm{X}_n)$. Moreover, due to the complexity of health economic decision models, after we obtain the samples of $\bm{\theta}$ from $p(\bm{\theta}|\bm{X}_n)$, the evaluation of $\mbox{NB}_d(\bm{\theta} )$ may also take a long computational time. \cite{ades2004expected} These computational challenges motivate the development of more efficient EVSI calculation methods, such as the original GA approach. \cite{jalal2018gaussian,heath2020calculating,jackson2021value}
\subsection{Approximating the Conditional Expectation of a Prior Parameter using the Gaussian Approximation Approach}
This section reviews the original GA approach. Assume we are aiming to approximate $\mathbb{E}_{\phi}[\phi|\bm{X}_n]$, where $\phi$ represents a univariate subset of the parameters$\bm{\theta}$ within the health economic model. The prior of $\phi$ is assumed to be Gaussian distributed with mean $\mu_0$ and variance $\frac{\sigma^2}{n_0}$:
\begin{equation}
\phi \sim N\left(\mu_0,\frac{\sigma^2}{n_0}\right), \label{eq:prior}
\end{equation}
where $n_0$ is called the prior effective sample size (ESS) representing the amount of information contained in $\phi$, and $\sigma^2$ is the individual-level variance of the proposed data collection. 
$n_0$ can be estimated by the methods proposed in the original GA paper or the nonparametric regression-based method recently proposed by Li et al. \cite{jalal2018gaussian,li2023}

Moreover, we denote the sample mean of the simulated dataset with sample size $n$ by $\widebar{\bm{X}}_n$. Assuming that, given $\phi$, each observation in the simulated dataset is also Gaussian distributed with mean $\phi$ and variance $\sigma^2$, the sample mean  $\widebar{\bm{X}}_n $ given $\phi$ is also Gaussian such that:
\begin{equation}
\widebar{\bm{X}}_n |\phi \sim N\left(\phi,\frac{\sigma^2}{n}\right), \label{eq:likelihood}
\end{equation}

Since both $\phi$ and $\widebar{\bm{X}}_n |\phi $ are Gaussian, we can integrate out $\phi$ and the marginal distribution of $\widebar{\bm{X}}_n$ is still Gaussian:

\begin{equation}
\widebar{\bm{X}}_n  \sim N\left(\mu_0, \frac{\sigma^2}{v n_0 }\right),  v =  \frac{n }{ n_0 + n}  \label{eq:marginal_mean}
\end{equation}

Because the prior distribution and likelihood function are conjugate, the conditional mean of $\phi$ given the simulated dataset $\bm{X}_n$ is a weighted sum between the sample mean and prior mean:
\begin{equation}
\mathbb{E}_{\phi}[\phi|\bm{X}_n] = (1 - v) \mu_0 + v \widebar{\bm{X}}_n , \label{eq:posteior_mean_eq}
\end{equation}
where $\mathbb{E}_{\phi}[\phi|\bm{X}_n]$ is a linear transformation of a Gaussian random variable $\widebar{\bm{X}}_n$ and thus follows a Gaussian distribution.  After simplification, the marginal distribution of $\mathbb{E}_{\phi}[\phi|\bm{X}_n] $ becomes
\begin{equation}
\mathbb{E}_{\phi}[\phi|\bm{X}_n] \sim N\left(\mu_0, \frac{v \sigma^2}{n_0 }\right). \label{eq:posterior}
\end{equation}

Since the prior distribution of $\phi$ is Gaussian with mean $\mu_0$ and variance $\frac{\sigma^2}{n_0}$, we 
can also rescale the samples of $\phi$ drawn from the prior $p(\phi)$ to obtain a Gaussian random variable with the same mean and variance as the distribution of $\mathbb{E}_{\phi}[\phi|\bm{X}_n]$. To be specific, for $\phi$ with the prior distribution $N\left(\mu_0,\frac{\sigma^2}{n_0}\right)$, we have 
\begin{equation}
\left ( \sqrt{\frac{ n }{n_0 + n  }}\left(\phi-\mu_0 \right) + \mu_0\right )\sim N\left(\mu_0, \frac{v \sigma^2}{n_0 }\right). \label{eq:post_mean}
\end{equation}

Following equation \ref{eq:post_mean}, we can construct the distribution of $\mathbb{E}_{\phi}[\phi|\bm{X}_n]$ by linearly transforming the samples drawn from the prior distribution $p(\phi)$. Let us denote $M$ samples of $\phi$ as $\phi^i \sim p(\phi),  i = 1, \dots, M$, the distribution of  $\mathbb{E}_{\phi}[\phi|\bm{X}_n]$ can be approximated by samples of $\mu_X$, which are defined as:
\begin{align}
\mu_{X}^i&= \sqrt{v} \phi^i +(1 - \sqrt{v} )\mu_0,\label{eq:GA_result}
\end{align}

Since in health economic evaluation, the prior distribution of $\phi$ is usually determined using previous studies, the shape of $\phi$ tends to be unimodal and concentrated around a specific value. Therefore, approximating the prior distribution of $\phi$ by a Gaussian distribution and updating the conditional expectation of $\phi$ by rescaling the samples drawn from $p(\phi)$ using equation \ref{eq:GA_result} is reasonable. Moreover, Jalal and Alarid-Escudero also demonstrated that the original GA approach can well approximate the marginal distribution of the conditional expectation of $\phi$ for most of the commonly used likelihood functions and prior distributions in the original GA paper.\cite{jalal2018gaussian} 

Jalal and Alarid-Escudero suggest that $\mathbb{E}_{\bm{\theta}}[ \mbox{NB}_d(\bm{\theta}) |{\bm{X}_n}]$, required for EVSI estimation, can be estimated by directly plugging the $\mathbb{E}_{\phi}[\phi|\bm{X}_n]$ approximated by the GA into a fitted regression model. This method is named as the `linear meta-model' in the original GA approach. However, since the original GA approach only uses the expectation (first moment) of $\phi$ to approximate $\mathbb{E}_{\bm{\theta}}[ \mbox{NB}_d(\bm{\theta}) |{\bm{X}_n}]$, the estimated $\mathbb{E}_{\bm{\theta}}[ \mbox{NB}_d(\bm{\theta}) |{\bm{X}_n}]$ will be biased if the net benefit function is nonlinear, which will affect the accuracy of the estimated EVSI. In the next section, we will discuss how to provide more accurate estimates of $\mathbb{E}_{\bm{\theta}}[ \mbox{NB}_d(\bm{\theta}) |{\bm{X}_n}]$ using the variance (second moment) of $\phi$.

\subsection{Taylor Series Expansions for the Conditional Expectation of Net Benefit}

Taylor series expansions can provide a more accurate estimation for $\mathbb{E}_{\bm{\theta}}[ \mbox{NB}_d(\bm{\theta}) |{\bm{X}_n}]$ by introducing the higher-order correction terms instead of only using the first-order term of $\phi$ as in the original GA.\cite{madan2014strategies} To demonstrate the use of Taylor series expansions, we begin by rewriting $\mathbb{E}_{\bm{\theta}}\left[\mbox{NB}_d(\bm{\theta}) |{\bm{X}_n} \right]$ using the law of total expectation:
\begin{align}
\mathbb{E}_{\bm{\theta}}\left[\mbox{NB}_d(\bm{\theta}) |\bm{X}_n \right] 
&= \mathbb{E}_{\phi}\left[\mathbb{E}_{\bm{\theta}}\left[\mbox{NB}_d(\bm{\theta}) |\phi \right] |\bm{X}_n\right]  \notag \\ &= \mathbb{E}_{\phi}\left[ g_d(\phi) |\bm{X}_n\right],\label{eq:EVPPI}
\end{align}
where $\mathbb{E}_{\bm{\theta}}\left[\mbox{NB}_d(\bm{\theta}) |\phi \right] $ can be viewed as a function of $\phi$ and denoted $g_d(\cdot)$.

To approximate $\mathbb{E}_{\bm{\theta}}[ \mbox{NB}_d(\bm{\theta}) |{\bm{X}_n}]$ using $\mathbb{E}_{\phi}[\phi|\bm{X}_n]$, we can apply second-order Taylor series expansions to reexpress $ \mathbb{E}_{\phi}\left[ g_d(\phi) |\bm{X}_n\right]$ as a function of  the conditional net benefit function $g_d(\cdot)$ and the conditional moments of $\phi$ :\cite{wolter2007introduction}
\begin{align}
\mathbb{E}_{\phi}\left[ g_d(\phi) |\bm{X}_n\right] &\approx g_d(\mathbb{E}_{\phi}[\phi|\bm{X}_n])  \notag  \\
& + \frac{\mbox{Var}_{\phi}[\phi|\bm{X}_n]}{2} g_d''(\mathbb{E}_{\phi}[\phi|\bm{X}_n]) \label{eq:Taylor_series},
\end{align}where $g_d''(\cdot)$ denotes the second-order derivative of $g_d(\cdot)$ with respect to the input value. 

Following equations \ref{eq:EVPPI} and \ref{eq:Taylor_series}, the estimation of $\mathbb{E}_{\bm{\theta}}[\mbox{NB}_d(\bm{\theta})|\bm{X}_n]$ can be decomposed into the estimation of $\mathbb{E}_{\phi}[\phi|\bm{X}_n]$, $\mbox{Var}_{\phi}[\phi|\bm{X}_n]$, $g_d(\cdot)$ and its second-order derivative $g_d''(\cdot)$. In the previous section, we have demonstrated that $\mathbb{E}_{\phi}[\phi|\bm{X}_n]$ can be estimated using the conventional GA approach. The estimation of  $\mbox{Var}_{\phi}[\phi|\bm{X}_n]$,  $g_d(\cdot)$, and $g_d''(\cdot)$ will be elaborated in the next two sections. After obtaining all of the approximated quantities, they can be combined together using equation \ref{eq:Taylor_series} to produce $\mathbb{E}_{\bm{\theta}}[\mbox{NB}_d(\bm{\theta})|\bm{X}_n]$ for EVSI estimation. 

Although equation \ref{eq:Taylor_series} utilizes a second-order Taylor series approximation for estimating $\mathbb{E}_{\bm{\theta}}[\mbox{NB}_d(\bm{\theta})|\bm{X}_n]$, it is possible to employ a higher-order Taylor series approximation to include additional correction terms of $\phi$ for approximating $\mathbb{E}_{\bm{\theta}}[\mbox{NB}_d(\bm{\theta})|\bm{X}_n]$. However, we do not recommend this approach for two reasons. Firstly, the second-order Taylor series approximation is generally sufficient for most scenarios because the influence of higher-order terms diminishes quickly as the dataset's sample size increases. Therefore, including more correction terms may not significantly improve the accuracy of the approximation. \cite{wolter2007introduction} Secondly, in the upcoming sections, we will utilize spline regression models to approximate the conditional net benefit function $g_d(\cdot)$ in equation \ref{eq:Taylor_series}. However, given the limited number of samples used for fitting, the spline may not accurately approximate the higher-order derivative of the net benefit function, which means employing such information in a Taylor series approximation may not necessarily improve the precision of the conditional net benefit estimates.    \cite{hastie1990generalized,wood2006generalized} Therefore, applying a higher-order Taylor series approximation is not required in our approach. 

Finally, note that this article limits the parameter $\phi$ to be univariate for ease of explanation but our methodology can be generalized to the multivariate case. The estimation method for $\mathbb{E}_{\bm{\theta}}[\mbox{NB}_d(\bm{\theta})|\bm{X}_n]$ when $\bm{\phi}$ is multivariate is introduced in the appendix.

\subsection{Approximating Conditional Variance of a Prior Parameter using Expected Fisher Information}

In this section, we introduce the methods that can be used to estimate the conditional variance $\mbox{Var}_{\phi}[\phi|\bm{X}_n]$. One way to estimate $\mbox{Var}_{\phi}[\phi|\bm{X}_n]$ is using the numerical integration method such as MCMC. Nevertheless, because estimating EVSI requires us to estimate the conditional variance for a large number of simulated datasets, this means we need to repeat MCMC many times and that will result in a high computational cost. 

Alternatively, if the Gaussian assumption of $\phi$ and $\bm{X}_n$ is satisfied, we can show that the conditional variance is a fixed constant and independent of the dataset, which implies  $\mbox{Var}_{\phi}[\phi|\bm{X}_n] = \mathbb{E}_{\bm{X}_n} [\mbox{Var}_{\phi}[\phi|\bm{X}_n]] $.  Then, using the iterated law of total variance, the conditional variance is equal to:
\begin{align}
\mbox{Var}_{\phi}[\phi|\bm{X}_n] 
&= \mathbb{E}_{\bm{X}_n} [\mbox{Var}_{\phi}[\phi|\bm{X}_n]] \notag \\
&= \mbox{Var}_{\phi}[\phi] - \mbox{Var}_{\bm{X}_n} [\mathbb{E}_{\phi}[\phi|\bm{X}_n]] \notag \\& =  \frac{v\sigma^2}{n_0}.  \label{eq:exp_posterior_var}
\end{align}

However, since the Gaussian assumption of $\phi$ is often not strictly satisfied, directly using $\frac{v\sigma^2}{n_0}$ to approximate the conditional variance may not be always accurate. Therefore, to balance the computational efficiency and estimation accuracy, we suggest using the expected Fisher information function to estimate $\mbox{Var}_{\phi}[\phi|\bm{X}_n]$.\cite{bernardo2009bayesian}   

The expected Fisher information is a crucial concept in statistical estimation theory. Utilizing asymptotic theory, it approximates $\mbox{Var}_{\phi}[\phi|\bm{X}_n]$ based on the estimates of $\phi$ obtained from the dataset. This information quantifies the amount of knowledge contained in the collected dataset, empowering us to draw meaningful inferences and make informed decisions based on the available information.

The expected Fisher information function is defined as the expectation of the second-order derivative of the log-likelihood function of $\phi$ over the dataset $\bm{X}_n$.\cite{van2000asymptotic} When samples included in the dataset with the sample size $n$ are independent and identically distributed, the expected Fisher information function can be written as:
\begin{equation}
 I_n(\phi) = -n \mathbb{E}_{X} \left[    \frac{d^2}{d\phi^2} \log\left(f(X|\phi)\right)     \right]. \label{eq:Var_results}
\end{equation}The functional forms of  $I_n(\cdot)$ for commonly used likelihood functions are included in the appendix. The functional form of  $I_n(\cdot)$ can also be approximated based on the numerical methods implemented in the \texttt{R} package \textbf{mle.tools}, but that will typically require more computational time. \cite{mazucheli2017mle}

Using the asymptotic properties of the conditional distribution of $\phi$, we can plug the conditional expectation  $\mathbb{E}_{\phi}[\phi|\bm{X}_n]$ into the reciprocal of the expected Fisher information function to approximate the corresponding conditional variance $\mbox{Var}_{\phi}[\phi|\bm{X}_n]$ by $I_n^{-1}(\mathbb{E}_{\phi}[\phi|\bm{X}_n])$. Since in equation \ref{eq:GA_result}, we have shown that the marginal distribution of $\mathbb{E}_{\phi}[\phi|\bm{X}_n]$ can be approximated by the samples  $\mu_{X}^1 , \dots, \mu_{X}^M$ based on the conventional GA, it means that we can evaluate $I_n^{-1}(\mu_{X}^i) , \, i=1, \dots, M$ and take $I_n^{-1}(\mu_{X}^i)$ as the conditional variance estimate associated with $\mu_{X}^i$. \cite{bernardo2009bayesian}

Lastly, since the asymptotic conditional variance approximated by the expected Fisher information is usually greater than the true conditional variance, \cite{strimmer2022statistical} we can further adjust the conditional variance provided by the expected Fisher information using the law of total variance. Because equation \ref{eq:exp_posterior_var} suggests that the average of the conditional variance should be equal to $\frac{v\sigma^2}{n_0}$, whilst $\frac{1}{M}\sum_{i = 1}^M I_n^{-1}(\mu_{X}^i)$ is not always equal to that, we can additionally multiply each of $I_n^{-1}(\mu_{X}^i)$ by a constant $C$ to make the averaged conditional variance of $\phi$ equal to $\frac{v\sigma^2}{n_0}$.  Such $C$ can be solved by evaluating:
\begin{equation}
C =\frac{\left(  \frac{v\sigma^2}{n_0} \right) }{\left( \frac{1}{M} \sum_{i = 1}^M I_n(\mu_{X}^i)\right)}.
\label{eq:adjusted_const}
\end{equation}
After adjustments, the conditional variance associated with $\mu_{X}^i , \, i=1, \dots, M$ becomes:

\begin{equation}
CI_n(\mu_{X}^i), \, i=1, \dots, M. \label{eq:Var_results_2}
\end{equation}

\subsection{Approximating Conditional Expectation of Net Benefit using Splines}

This section introduces how to approximate the functional form of  $\mathbb{E}_{\bm{\theta}}\left[\mbox{NB}_d(\bm{\theta}) | \phi \right]$ using splines and the PA dataset. The marginal distribution of $\mathbb{E}_{\bm{\theta}}\left[\mbox{NB}_d(\bm{\theta}) | \bm{X}_n \right]$ can then be approximated based on the estimated $\mathbb{E}_{\phi}[\phi|\bm{X}_n]$, $\mbox{Var}_{\phi}[\phi|\bm{X}_n]$ and the fitted spline. 

Splines are a type of flexible regression model that can characterize the nonlinear relationship between the responses and predictors using a series of basis functions. \cite{kharroubi2011estimating,wood2006generalized}  As Strong et al. introduced,\cite{strong2015estimating} using the PA datasets, we can regress the samples of net benefits given by the decision $d$ on the samples of the parameter of interest, $\phi$, to obtain the fitted spline $\hat g_d(\phi)$. $\hat g_d(\phi)$ is an approximation of the function form of $\mathbb{E}_{\bm{\theta}}\left[\mbox{NB}_d(\bm{\theta}) | \phi \right]$ and reflects how much economic benefit can be obtained by eliminating all the uncertainty around the parameter of interest. This approach is computationally efficient and can accurately approximate the complex net benefit function, provided a sufficient number of probability analysis samples are available. \cite{heath2020calculating} For a detailed introduction to the implementation and model diagnostics of using splines to approximate the expected net benefit conditioning on the specific parameters, see Strong et al.\cite{strong2014estimating}

After the functional form of $\mathbb{E}_{\bm{\theta}}\left[\mbox{NB}_d(\bm{\theta}) | \phi \right] $ is approximated by the fitted spline $\hat g_d(\phi)$, the second-order derivative of $\mathbb{E}_{\bm{\theta}}\left[\mbox{NB}_d(\bm{\theta}) | \phi \right]$ can be naturally approximated by the second-order derivative of the fitted splines, $\hat g_d''(\phi)$.  $\hat g_d''(\phi)$ is determined by the second-order derivative of the basis functions included in the fitted model and can be efficiently computed using the \texttt{R} package \textbf{splines2}. \cite{wang2021package,wang2021shape}\footnote{In this article, the type of the spline used in implementation is constructed by B-spline basis functions, but other types of basis functions such as natural spline, C-splines, and I-splines basis functions can also be considered. Additionally, to ensure sufficient flexibility of the spline, we generally recommend that the basis function include up to a third degree of derivative. The implementation of the spline is based on the R package \textbf{splines2}, which is implemented based on Rcpp, C++ and has high computational efficiency. Moreover, \textbf{splines2} can efficiently compute the first and second derivative of a set of splines, which makes it a good fit for the spline-based Taylor series expansion framework.\cite{wang2021package,wang2021shape}}

In equation \ref{eq:Taylor_series}, the conditional expectation of the net benefit can then be approximated using the fitted splines:
\begin{align}
\mathbb{E}_{\bm{\theta}}[\mbox{NB}_d(\bm{\theta})|\bm{X}_n] &\approx \hat g_d(\mathbb{E}_{\phi}[\phi|\bm{X}_n])  \notag  \\
& + \frac{\mbox{Var}_{\phi}[\phi|\bm{X}_n]}{2} \hat g_d''(\mathbb{E}_{\phi}[\phi|\bm{X}_n]). 
\label{eq:Taylor_series_spline}
\end{align}

Moreover, using equations \ref{eq:GA_result} and \ref{eq:Var_results_2}, the marginal distribution of $\mathbb{E}_{\bm{\theta}}[\mbox{NB}_d(\bm{\theta})|\bm{X}_n]$ can be approximated by the samples of the estimated conditional net benefit, $\mathbb{E}_{\bm{\theta}}[\mbox{NB}_d(\bm{\theta})|\bm{X}_n^i],\, i=1, \dots, M$, which are defined as:
\begin{align}
\mathbb{E}_{\bm{\theta}}[\mbox{NB}_d(\bm{\theta})|\bm{X}_n^i] &\approx \hat  g_d(\mu_{X}^i) \notag  \\& + \frac{CI_n^{-1}(\mu_{X}^i)}{2}\hat  g_d''(\mu_{X}^i).
\label{eq:Taylor_series_spline}
\end{align}

\subsection{EVSI calculation}
Finally, we can repeat the above procedures to approximate the marginal distribution of $\mathbb{E}_{\bm{\theta}}[\mbox{NB}_d(\bm{\theta})|\bm{X}_n]$ using samples $\mathbb{E}_{\bm{\theta}}[\mbox{NB}_d(\bm{\theta})|\bm{X}_n^i]$ for all decisions $d = 1, \dots, D$. Using equation \ref{eq:EVSI_fixed}, EVSI for a study design with the sample size $n$ can be estimated by the samples $\mathbb{E}_{\bm{\theta}}[\mbox{NB}_d(\bm{\theta})|\bm{X}_n^i]$ such that:
\begin{align}
\text{EVSI}(n) &\approx \frac{1}{M} \sum_{i = 1}^M \left( \max_d \mathbb{E}_{\bm{\theta}}[\mbox{NB}_d(\bm{\theta})|\bm{X}_n^i]  \right) \notag \\&- \max_d \frac{1}{M} \sum_{i = 1}^M \mathbb{E}_{\bm{\theta}}[\mbox{NB}_d(\bm{\theta})|\bm{X}_n^i] . \label{eq:app_EVSI}
\end{align}

The algorithm for estimating EVSI for $k$ different sample sizes $n_1, \dots, n_k$  using spline-based Taylor series expansions and GA is summarized in algorithm \ref{alg:GA_Taylor}. Once the prior effective sample size $n_0$, the fitted splines $\hat g_d(\cdot)$ and the expected Fisher information function $I_n(\cdot)$ are determined, the proposed algorithm can estimate EVSI across different sample sizes with low computational cost. Therefore, the proposed approach is more computationally efficient than nonparametric regression-based approaches and methods based on advanced Monte Carlo methods for identifying optimal study designs that maximize EVSI.\cite{heath2020calculating,strong2015estimating,brennan2007efficient,kharroubi2011estimating,menzies2016efficient,hironaka2020multilevel,fairley2020optimal} 
\begin{algorithm}
\caption{Estimating EVSI using GA and Spline-Based
Taylor Series Expansions}\label{alg:GA_Taylor}
 \SetKwInOut{Input}{Inputs}
 \Input{A PA dataset of size $M$; The likelihood function of the parameter of interest $\phi$; The set $\{n_1, \dots, n_k\}$ which includes $k$ distinct sample sizes required for EVSI estimation;}
 Estimating $n_0$ for $\phi$ using the method proposed by Li et al. or methods included in Jalal and Alarid-Escudero's article \cite{jalal2018gaussian,li2023} \ ;
 Estimating the functional form for the conditional net benefit of each decision by regressing the net benefit samples on the samples of $\phi$ using the spline. The obtained fitted splines are $\hat g_d(\phi), d = 1, \dots, D$\;
 Finding the expected Fisher information function $I_n(\cdot)$ based on the likelihood function of $\phi$ analytically or numerically \;
  \For{$n \in \{n_1, \dots, n_k\}$ }{
    Compute $\mu_{X}^i = \sqrt{v} \phi^i +(1 - \sqrt{v} )\mu_0$ for $i = 1, \cdots, M$, where $v =  \frac{n}{n + n_0}$ \;
    Compute $C = \left( \frac{v\sigma^2}{n_0} \right) /\left( \frac{1}{M} \sum_{i = 1}^M I_n(\mu_{X}^i)\right)$\footnotemark  \;
    Compute the adjusted conditional variance $CI_n(\mu_{X}^i)$ for $i = 1, \cdots, M$\;
    \For{$d= 1$ \KwTo $D$}{
        Compute the adjusted conditional net benefit $\mathbb{E}_{\bm{\theta}}[\mbox{NB}_d(\bm{\theta})|\bm{X}_n^i]  = g_d(\mu_{X}^i)  + \frac{CI_n^{-1}(\mu_{X}^i)}{2} g_d''(\mu_{X}^i),$  for $i=1, \dots, M.$
        }
     $\widehat{\text{EVSI}}(n) = \frac{1}{M} \sum_{i = 1}^M \left( \max_d \mathbb{E}_{\bm{\theta}}[\mbox{NB}_d(\bm{\theta})|\bm{X}_n^i] \right) - \max_d \frac{1}{M} \sum_{i = 1}^M \mathbb{E}_{\bm{\theta}}[\mbox{NB}_d(\bm{\theta})|\bm{X}_n^i] $
  }

 \KwResult{The set of the estimated EVSI $\{\widehat{\text{EVSI}}(n_1), \dots, \widehat{\text{EVSI}}(n_k)\}$}
\end{algorithm}
\footnotetext{The step $6$ is optional. Without this step, the user can replace $CI_n(\phi_i)$ with $I_n(\phi_i)$ in the rest of this algorithm. Incorporating the constant $C$ into the algorithm can enhance the accuracy of the EVSI estimation especially when the sample size of the proposed study is small, but it also adds complexity to the implementation process.”}

\section*{SIMULATION STUDY}

\subsection{Case study I: Gaussian parameters and nonlinear net benefit functions}

In the first case study, our augmented GA method, based on splines and Taylor series expansions (TGA) is used to evaluate EVSI for four stylized examples with Gaussian distributed parameters and nonlinear net benefit functions. EVSI estimates from our TGA method are compared to the conventional GA approach and the nonparametric regression-based method to demonstrate accuracy. Additionally, analytic methods are used to simplify the nested Monte Carlo EVSI estimator to be used as the comparator in all four examples.

\subsubsection{Incremental net benefit function: }

Our decision problem compares two potential interventions. To simplify the calculation, we can derive the incremental net benefit function using the net benefit functions of two decision options by subtracting one net benefit function from the other, i.e., $\mbox{INB}(\bm{\theta}) = \mbox{NB}_1(\bm{\theta}) - \mbox{NB}_2(\bm{\theta})$. 
We test the robustness of TGA by specifying different functional forms for $\mbox{INB}(\bm{\theta})$ in four hypothetical scenarios, including both linear and nonlinear scenarios. The incremental net benefit functions are summarized in Table \ref{tab:case_1}.

Using the incremental net benefit function, EVSI can be calculated using the conditional expectation of $\mbox{INB}(\bm{\theta})$ given the simulated dataset:\footnote{The equivalence of the EVSI, whether expressed through incremental net benefit functions or net benefit functions, is detailed in Section E of the appendix.} \cite{heath2019estimating}

\begin{align}
    \text{EVSI}(n) &= \mathbb{E}_{\bm{X}_n} \left [ \max \left \{ \mathbb{E}_{\bm{\theta}}[ \mbox{INB}(\bm{\theta}) |{\bm{X}_n}], 0 \right \} \right ]\notag \\ 
    & -  \max \left \{\mathbb{E}_{\bm{X}_n} \left[
  \mathbb{E}_{\bm{\theta}}[ \mbox{INB}(\bm{\theta}) |\bm{X}_n] \right], 0 \right \}. \label{eq:EVSI_INB_2}
\end{align}

Therefore, we can approximate the distribution of $\mathbb{E}_{\bm{\theta}}[ \mbox{INB}(\bm{\theta}) |{\bm{X}_n}]$ and compare it with $0$ to estimate EVSI. 

\begin{table}[t]
\small\sf\centering
\begin{tabular}{cc}
\toprule
Scenario &Incremental Net Benefit Function\\
\midrule
\texttt{1}& $\mbox{INB}(\theta) = -100 + 5000\theta$\\
\texttt{2} &$\mbox{INB}(\theta) = -1000 + 5000\theta^2 $\\
\texttt{3} &$\mbox{INB}(\theta) = -500 + 5000 \theta^4$\\
\texttt{4} &$\mbox{INB}(\theta_1, \theta_2) = -1500 + 5000\theta_1^2 + 5000 \theta_2^4 $\\
\bottomrule
\end{tabular}
\caption{Incremental net benefit functions for four stylized studies in case study I.\label{tab:case_1}}
\end{table}

\subsubsection{Parameter of interest and dataset generation:}

For the first three case studies where the parameter of interest $\theta$ is univariate, the prior distribution of $\theta$ is set to be Gaussian with mean $\mu_0 = 0$, $\sigma^2 = 1$ and $n_0 = 5$. Consequently, the variance of the Gaussian distribution is  $\frac{\sigma^2}{n_0} = \frac{1}{5}$. This selection reflects a level of uncertainty in $\bm{\theta}$ that is representative of typical scenarios in real-world health economic evaluations.

For the first three case studies where the parameter of interest $\theta$ is univariate, the prior distribution of $\theta$ is set to be Gaussian with mean $\mu_0 = 0$ and variance $\frac{\sigma^2}{n_0} = \frac{1}{5}$:
\begin{equation}
\theta \sim N \left (0, \frac{1}{5} \right ).
\end{equation}The likelihood function of $\theta$ is also Gaussian distributed for the first three scenarios. $n$ collected data are drawn from a Gaussian likelihood function such that:
\begin{equation}
 X_i |\theta \sim N \left (\theta, 1 \right ), i = 1,\dots,n. \label{eq:normal_lik}
\end{equation}
The prior distribution for the fourth bivariate incremental net function is an independent bivariate normal distribution with a mean of 0 and a variance of  $\frac{1}{5}$. The likelihood function of $(\theta_1, \theta_2)$ is also modelled as an independent bivariate Gaussian distribution with a variance $1$.\footnote{In the first case study, the terms $\bm{\theta}$ and $\phi$ are used interchangeably, as $\phi$ represents the entire parameter set in this specific context. }

\subsubsection{Method 1 -- Analytic method:}
For each data collection exercise, we draw $M = 10^5$ samples from the  prior distribution $p(\bm{\theta})$ and obtain samples $\bm{\theta}^1, \dots, \bm{\theta}^M$.   Next, we generate a simulated dataset for each sample of $\bm{\theta}$, and achieve  $\bm{X}_n^1, \dots, \bm{X}_n^M$. Due to the conjugacy, the conditional expectation of the incremental net benefit for each simulated dataset,  $\mathbb{E}_{\bm{\theta}}[\mbox{INB}(\bm{\theta})|\bm{X}_n^i]$, can be analytically computed for each simulated dataset. The analytic solutions of $\mathbb{E}_{\bm{\theta}}[\mbox{INB}(\bm{\theta})|\bm{X}_n^i]$ for the four case studies are summarized in the appendix.

Using the simulated dataset with the sample size $n$ and  equation  \ref{eq:EVSI_INB_2}, we can estimate EVSI for different $ \mbox{INB}(\bm{\theta})$ functions by:
\begin{align}
\widehat{ \mbox{EVSI}}(n) &= \frac{1}{M} \sum_{i = 1}^M  \max \left \{ \mathbb{E}_{\bm{\theta}}[\mbox{INB}(\bm{\theta})|\bm{X}_n^i], 0  \right \}  \notag \\ \qquad &-\max \left \{  \frac{1}{M} \sum_{i = 1}^M \mathbb{E}_{\bm{\theta}}[\mbox{INB}(\bm{\theta})|\bm{X}_n^i], 0  \right \}.\label{eq:EVSI_case1}
\end{align}

\subsubsection{Method 2 -- Nonparametric regression-based method:}

We generate $M=10^5$ samples of the dataset $\bm{X}_n$ and use the sample mean $\widebar{\bm{X}}_n$ as the summary statistic for each simulated dataset.  We regress the incremental net benefit samples on all the samples of $\widebar{\bm{X}}_n$ using splines to estimate $E_{\bm{\theta}}[\mbox{INB}(\bm{\theta})|\bm{X}_n]$. The fitted values are then extracted from the regression models and used to estimate $E_{\bm{\theta}}[\mbox{INB}(\bm{\theta})|\bm{X}_n]$.

\subsubsection{Method 3 -- Linear meta-modelling Gaussian approximation:}

Since the Gaussian assumption is strictly satisfied, we can derive values of prior effective sample sizes $n_0$ analytically. For the first three $\mbox{INB}(\bm{\theta})$ functions, the prior effective sample size $n_0$ is $5$. For the fourth incremental net benefit function, the effective sample sizes for $\bm{\theta} = ({\theta_1, \theta_2})$ are both $5$. The linear meta-model used to produce $E_{\bm{\theta}}[\mbox{INB}(\bm{\theta})|\bm{X}_n]$ estimates is constructed by regressing the $10^5$ incremental net benefit samples on the samples of $\bm{\theta}$ drawn from the prior distribution using splines.

\subsubsection{Method 4 -- Spline-Based Taylor series Gaussian approximation:}

 The prior effective sample sizes $n_0$ of TGA are the same as the conventional GA. Since both the likelihood function and prior distribution are Gaussian, the closed-form solution of $\mbox{Var}_{\phi}[\phi|\bm{X}_n]$ can be derived. For each $\mbox{INB}(\bm{\theta})$, we approximate its function form using a spline which is fitted by $M = 10^5$ incremental net benefit and $\bm{\theta}$ samples. We can then generate the samples of $E_{\bm{\theta}}[\mbox{INB}(\bm{\theta})|\bm{X}_n]$ via equation \ref{eq:Taylor_series_spline} and use them to approximate EVSI.

\subsection{Case study II: Calculating EVSI in a Markov Model}

In the second case study, we compare the accuracy of EVSI  given by the nonparametric regression-based method, conventional GA, and TGA using a Markov model included in Jalal and Alarid-Escudero. \cite{jalal2018gaussian} Four different data collection processes are considered to reduce the uncertainty in this Markov model and the corresponding EVSI are estimated. EVSI for each of the data collection processes is also computed using the nested Monte Carlo method for comparison. \cite{jalal2018gaussian}

\subsubsection{Incremental net benefit function: }

A Markov model consisting of three states is utilized to simulate a cohort with a hypothetical genetic disorder. Three interventions, $A$, $B$, and $C$, are available to prevent the progression of permanent disability resulting from the disorder, with $C$ being standard care. The failure of any of these interventions results in a reduction of patients' quality of life (QoL). This model has $15$ parameters in total, four of which are uncertain: mean hospital visits for interventions $A$ and $B$ (which are denoted by $\mu_{A}$ and $\mu_{B}$), and the failure probabilities for interventions $A$ and $B$ (which are denoted by $P_{A}$ and $P_{B}$). The conditional net benefit is linearly related to hospital visits for interventions $A$ and $B$, while the relationship between net benefit and failure probabilities of interventions $A$ and $B$ is nonlinear. For a full description of the model, see Jalal and Alarid-Escudero. \cite{jalal2018gaussian} 

\subsubsection{Parameter of interest and dataset generation:}

For this case study, we consider the four different data collection exercises that are included in  Jalal and Alarid-Escudero's work. These four data collection exercises aim to reduce the uncertainty in $\mu_{A}, \mu_{B}, P_{A}$ and $P_{B}$ respectively. The detailed information of the prior and likelihood for these four data collection processes are included in table \ref{tab:case_2}.

\begin{table*}[t]
\small\sf\centering
\begin{tabular}{ccc}
\toprule
Scenario &Prior Distribution & Likelihood Function\\
\midrule
\texttt{1}& $\mu_{A} \sim Gamma(\alpha = 10, \beta = 10)$ &$X_i \sim Poisson(\mu_A), i = 1, \dots, n$\\
\texttt{2} &$\mu_{B} \sim Gamma(\alpha = 20, \beta = 10)$ &$X_i \sim Poisson(\mu_B), i = 1, \dots, n$\\
\texttt{3} &$P_{A} \sim Beta(\alpha = 2, \beta = 8)$ &$X_i \sim Bernoulli(P_{A}), i = 1, \dots, n$\\
\texttt{4} &$P_{B} \sim Beta(\alpha = 3, \beta = 7)$&$X_i \sim Bernoulli(P_{B}), i = 1, \dots, n$\\
\bottomrule
\end{tabular}
\caption{Prior distribution and likelihood functions for the Markov Model in case study II. $\mu_{A}$ and $\mu_{B}$ are used to indicate the average number of hospital visits associated with interventions A and B. $P_{A}$  and $P_{B}$ represent the failure probabilities for interventions A and B.\label{tab:case_2}}
\end{table*}

\subsubsection{Method 1 -- Nested Monte Carlo method:}

The EVSI estimates from the nested Monte Carlo method are taken directly from Jalal and Alarid-Escudero's paper \cite{jalal2018gaussian}

\subsubsection{Method 2 -- Nonparametric regression-based method:}

We generate $M = 10^4$ samples of the parameter $\bm{\theta}$ from the prior $p(\bm{\theta})$, run the Markov model and compute the incremental net benefit samples. For all four data collection exercises, the summary statistics are the sample mean of the simulated dataset with the sample size $n$, $\widebar{\bm{X}}_n$. The increment net benefit samples are then regressed on the samples of $\widebar{\bm{X}}_n$ using splines.

\subsubsection{Method 3 -- Linear meta-modelling Gaussian approximation:}
We can derive that $n_0 = 10$ for all four data collection exercises as the likelihood function and prior distributions are conjugate. For each data collection exercise, the linear meta-model used to generate $E_{\bm{\theta}}[\mbox{INB}(\bm{\theta})|\bm{X}_n]$ estimates is the spline which is fitted by $M = 10^4$ incremental net benefit and parameter samples.

\subsubsection{Method 4 -- Spline-Based Taylor series Gaussian approximation:}

 Like the conventional GA approach,  $n_0 = 10$ for all data collection exercises.  For each data collection study, we approximate the function form of $\mathbb{E}_{\bm{\theta}}\left[\mbox{NB}_t(\bm{\theta}) |\phi \right] $ by regressing the $10^4$ incremental net benefit samples on the samples of parameters using  splines. The expected Fisher information functions for each scenario are derived analytically.

\section{RESULTS}

\subsection{Case study I: Gaussian parameters and nonlinear net benefit functions}

\begin{figure*}
\setlength{\fboxsep}{0pt}%
\setlength{\fboxrule}{0pt}%
\begin{center}
\centering
\includegraphics[width=.95\textwidth]{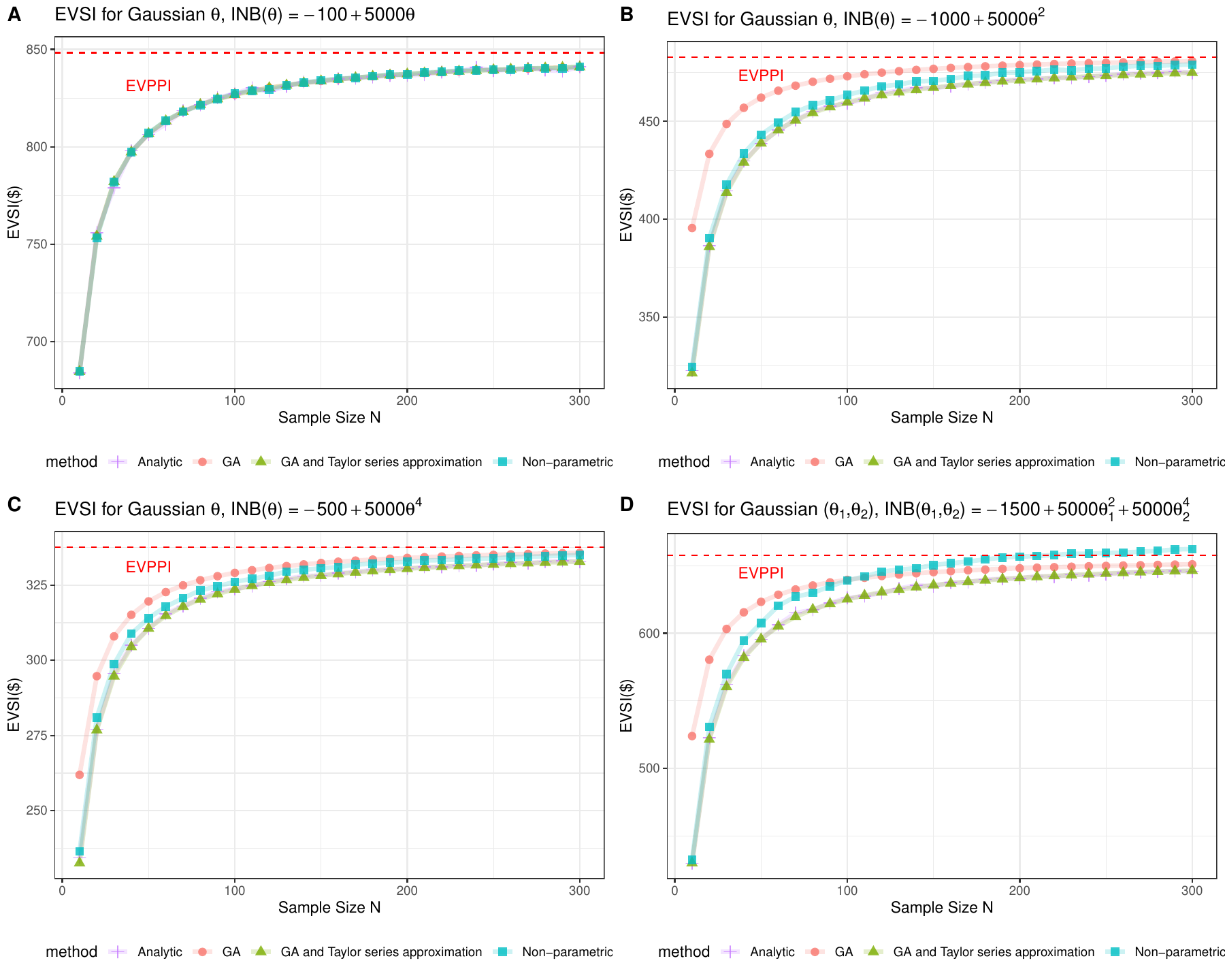} 
\end{center}
\caption{The Expected Value of Sample Information (EVSI) computed by analytic method (Analytic), conventional Gaussian Approximation (GA),  Spline-based Taylor series expansions and Gaussian Approximation  (GA and Taylor series approximation), nonparametric regression-based method (Non-nonparametric) for linear and nonlinear incremental net benefit functions with Gaussian distributed parameters. The expected value of  perfect information (EVPPI) is shown with the horizontal dashed lines.\label{fig:case1} }
\end{figure*}

Figure \ref{fig:case1} compares EVSI of four stylized net benefit functions with Gaussian distributed parameters computed using the analytic method,  the conventional GA approach, TGA, and the nonparametric regression-based method for different sample sizes  (between $10$ and $300$ in increments of $10$).   The expected value of perfect information (EVPPI)\cite{strong2014estimating} is computed by simulations and shown with the horizontal dashed line for each data collection exercise. EVPPI can quantify the economic benefit obtained by removing all the uncertainty around the parameter of interest and is the upper bound of EVSI. 

In subplot A of Figure \ref{fig:case1} , when $\mbox{INB}(\bm{\theta})$ is univariate linear, all four methods can accurately estimate EVSI. However, when $\mbox{INB}(\theta)$ is nonlinear, we can observe that GA largely overestimates EVSI. This is despite the Gaussian assumption being satisfied. Also, the nonparametric regression-based method may not accurately capture the correct functional form of $\mbox{INB}(\bm{\theta})$ when $\mbox{INB}(\bm{\theta})$ is highly nonlinear, thus it also overestimates EVSI for the third and fourth $\mbox{INB}(\bm{\theta})$ functions. Only EVSI curves given by TGA overlapped with the analytic method for all $\mbox{INB}(\bm{\theta})$, which suggests that when the Gaussian assumption is strictly satisfied, the TGA method can accurately estimate EVSI even if  $\mbox{INB}(\bm{\theta})$ is highly nonlinear.

\subsection{Case study II: Calculating EVSI in a Markov Model}

\begin{figure*}
\setlength{\fboxsep}{0pt}%
\setlength{\fboxrule}{0pt}%
\begin{center}
\centering
\includegraphics[width=.95\textwidth]{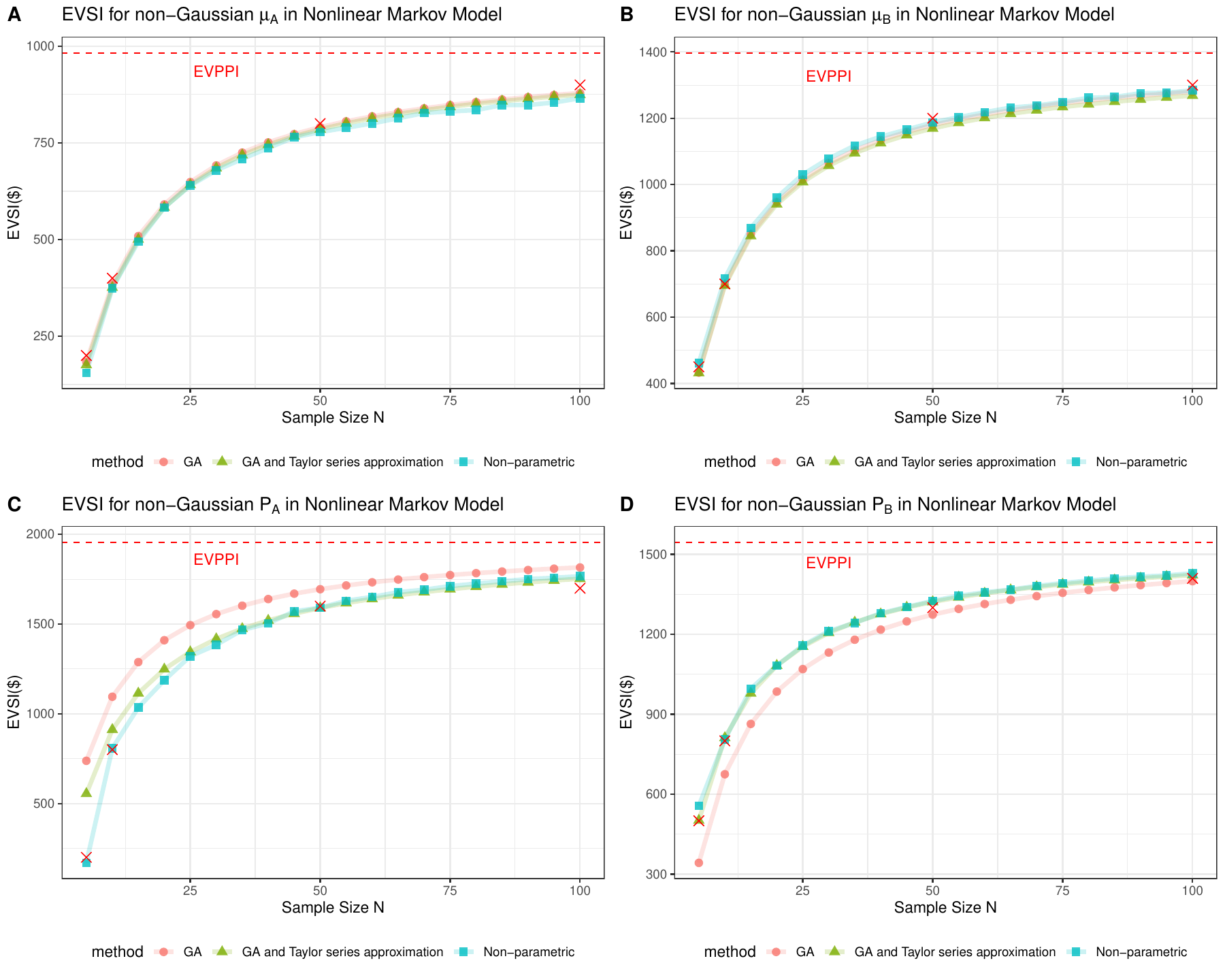} 
\end{center}
\caption{The Expected Value of Sample Information (EVSI) computed by conventional Gaussian Approximation (GA),  Spline-based Taylor series expansions and Gaussian Approximation  (GA and Taylor series approximation), nonparametric regression-based method (nonparametric) for a Markov Model across different sample sizes. EVSI estimated by standard nested Monte Carlo is denoted by the red cross. The expected value of partial perfect information (EVPPI) is shown with the horizontal dashed lines.\label{fig:case2} }
\end{figure*}

Figure \ref{fig:case2} compares EVSI of the Markov models computed by the conventional GA approach, TGA approach, and nonparametric regression-based method for different sample sizes  (between $5$ and $100$ in increments of $5$). EVSIs estimated by the nested Monte Carlo method for sample sizes equal to $5, 10, 50$, and $100$ are taken from Jalal and Alarid-Escudero and denoted by the red cross.  The EVPPI estimated by the nonparametric regression-based method proposed by Strong et al. is shown with the horizontal dashed lines for each data collection exercise. \cite{strong2014estimating}

When the relationship between conditional net benefits and parameters is nearly linear (subplots $A$ and $B$), the four approaches produce similar EVSI estimates. However, if the relationship is nonlinear (subplots $C$ and $D$), conventional GA may underestimate or overestimate EVSI compared to the other methods.  By contrast, when sample sizes are greater than $30$, TGA produces EVSI estimates that are similar to those generated by nested Monte Carlo and nonparametric regression-based methods. TGA's accuracy is lower than that of nonparametric regression-based methods when sample sizes are less than $30$. This is because when the sample size of the simulated dataset is small, the accuracy of the conditional net benefit estimated by Taylor series expansions, the conditional mean of the parameter estimated by the Gaussian approximation, and the conditional variance of the parameter estimated by the Fisher information decreases.

\section{DISCUSSION}

This article presents a new algorithm, Spline-based Taylor series approximation and Gaussian approximation (TGA) for estimating EVSI.  In the TGA method, we estimate EVSI by approximating the conditional expectation of net benefits using $2$ steps. First, we use Taylor series expansion to approximate the conditional expectation of net benefits through the net benefit function and the conditional mean and variance of parameters. Subsequently, the net benefit function is approximated by the spline fitted to the PA dataset, and the conditional moments of the parameters are approximated by the conventional GA and expected Fisher information.

\subsection{Strengths and Limitations}

The TGA algorithm has several advantages over alternative EVSI estimation methods. Firstly, once the prior effective sample size is estimated, TGA can estimate EVSI across multiple sample sizes with minimal computational cost. This is more efficient than EVSI estimation algorithms where EVSI must be estimated separately for each sample size, i.e., their computational time scales linearly with the number of sample sizes. Methods with linear scaling include the nonparametric regression-based method and other estimation algorithms based on advanced Monte Carlo methods. \cite{strong2015estimating,brennan2007efficient,kharroubi2011estimating,menzies2016efficient,hironaka2020multilevel,heath2019estimating} Additionally, EVSI estimates obtained using TGA are smooth with respect to the sample sizes and convenient for determining study designs that maximize economic benefit through numerical optimization.\cite{fairley2020optimal} Finally, EVSI estimates from TGA are more accurate than conventional GA, especially when the net benefit function is highly nonlinear.\cite{heath2018efficient,heath2019estimating,jalal2018gaussian}

However, TGA's efficiency and accuracy may be affected in certain scenarios.  Firstly, if the parameters of interest have a high dimension and complex interactions, a spline with a lot of interaction terms may be required to accurately approximate the function form of the conditional net benefit function. As a result, more computational resources are required for computing the second-order derivative of the net benefit samples and this may reduce the efficiency of TGA. In this case, we can consider implementing the TGA methods using other nonparametric regression methods that are less impacted by `the curse of dimensionality', e.g., Artificial Neural Network, to approximate the functional form of the conditional net benefit function. A future study might examine the efficacy of these nonparametric regression models in estimating EVSI, particularly when the number of the parameters of interest is large. \cite{heath2016estimating,gelman2013bayesian}

Secondly, while the closed-form solution for the expected Fisher information is generally available for most data generating processes, there are some complex settings where conceptualizing and evaluating the expected Fisher information function can be challenging, e.g., if the likelihood function does not have a closed-form solution or it is difficult to identify the likelihood function of the data generating process. In such scenarios, alternative approaches to estimating EVSI, such as nonparametric regression-based methods or moment matching,\cite{strong2015estimating,heath2018efficient,heath2019estimating} may be more suitable.  

Lastly, because the approximation of the conditional variance of the parameters of interest based on the expected Fisher information is more accurate when the sample size of the design is relatively large, EVSI provided by TGA is less accurate when the sample size of the design is very small. The nonparametric regression-based method may be preferred over TGA in that scenario.

\subsection{Conclusion}

We introduced a novel EVSI estimation method that combines Taylor series approximation and GA. As shown by the two case studies, the proposed algorithm can efficiently estimate EVSI for multiple sample sizes, and is more accurate than conventional GA when the net benefit function is highly non-linear. We believe that our method could aid in the evaluation and optimization of study designs using EVSI, particularly when the underlying health economic decision  model is complex and includes a nonlinear structure.

\begin{acks}
Placeholder.
\end{acks}

\begin{sm}
Supplementary material for this article is available on the Medical Decision Making Web site at http://journals.sagepub.com/home.mdm
\end{sm}

\bibliographystyle{SageV}
\bibliography{MDM_ref.bib}

\end{document}